\documentclass[a4paper,10pt]{article}
\usepackage{amsmath,amssymb,color,appendix}
\usepackage{slashed}
\usepackage{bbm}
\usepackage{color}
\usepackage{chngpage}
\usepackage{graphicx}
\usepackage{dsfont}
\usepackage{upgreek}
\usepackage{color}
\usepackage{soul}
\usepackage{cancel}
\usepackage[normalem]{ulem}
\usepackage{float}
\definecolor{gre}{rgb}{0,0.4,0.3}

\newcommand{\be}{\begin{equation}}
\newcommand{\eqb}{\begin{eqnarray}}
\newcommand{\eqf}{\end{eqnarray}}
\newcommand{\bb}{\begin{equation}}
\newcommand{\ee}{\end{equation}}
\newcommand{\beq}{\begin{equation}}
\newcommand{\eeq}{\end{equation}}
\newcommand{\bea}{\begin{eqnarray}}
\newcommand{\eea}{\end{eqnarray}}

\definecolor{gre}{rgb}{0,0.4,0.3}
\usepackage{chngpage}

\usepackage{color}

\def\lbldef#1#2{\expandafter\gdef\csname #1\endcsname {#2}}

\def\href#1#2{#2}
\def\half{{1 \over 2}}

\newcommand{\ber}{\begin{eqnarray}}
\newcommand{\eer}{\end{eqnarray}}

\newcommand{\beqar}{\begin{eqnarray}}

\newcommand{\eeqar}{\end{eqnarray}}

\newcommand{\eeqarr}{\end{eqnarray}}
\newcommand{\ZZ}{{\rm \kern 0.275em Z \kern -0.92em Z}\;}

\def\CC{{\mathchoice
{\rm C\mkern-8mu\vrule height1.45ex depth-.05ex
width.05em\mkern9mu\kern-.05em}
{\rm C\mkern-8mu\vrule height1.45ex depth-.05ex
width.05em\mkern9mu\kern-.05em}
{\rm C\mkern-8mu\vrule height1ex depth-.07ex
width.035em\mkern9mu\kern-.035em}
{\rm C\mkern-8mu\vrule height.65ex depth-.1ex
width.025em\mkern8mu\kern-.025em}}}

\def\RR{{\rm I\kern-1.6pt {\rm R}}}

\def\ZZ{{\rm Z}\kern-3.8pt {\rm Z} \kern2pt}
\def\IB{\relax{\rm I\kern-.18em B}}
\def\ID{\relax{\rm I\kern-.18em D}}
\def\II{\relax{\rm I\kern-.18em I}}
\def\IP{\relax{\rm I\kern-.18em P}}

\newcommand{\bear}{\begin{eqnarray}}
\newcommand{\eear}{\end{eqnarray}}

\def\to{\rightarrow}

\def\to{\rightarrow}

\def\i{\iota}

\def\6{\partial}

\newfont{\namefont}{cmr10}
\newfont{\addfont}{cmti7 scaled 1440}
\newfont{\boldmathfont}{cmbx10}
\newfont{\headfontb}{cmbx10 scaled 1728}
\title{${\cal N}=2$ SUSY  Abelian Higgs model with hidden sector and BPS equations}
\author{Paola Arias$^{\,a}$ , Edwin Ireson$^{\,b}$,
Carlos N\'u\~nez$^{\,b }$\footnote{Also at CP3 Origins, SDU. Odense, Denmark.}\,
and Fidel Schaposnik$^{\,c}$  \\
~\\
{$^a\,$\it Departamento de $F\acute{\i}sica$, Universidad de Santiago de Chile} \\{\it Casilla 307, Santiago, Chile}
\\
~\vspace{-3 mm}
\\
{$^b\,$\it Department of Physics, Swansea University}\\
{\it Singleton Park, Swansea, SA2 8PP, UK}
\\
~\vspace{-3 mm}
\\
{$^c\,$\it Departamento de $F\acute{\i}sica$, Universidad Nacional de La Plata/IFLP} \\{\it CC 67, 1900 La Plata, Argentina}
}

\begin{document}
\maketitle
{\bf Abstract:} In this paper we study
a system  inspired on certain SUSY breaking models
and on more recent Dark Matter
scenarios. In our set-up, two Abelian gauge
fields interact via an operator that mixes their
kinetic terms. We find the extended Supersymmetric version of this system, that also generates
a Higgs portal type of interaction.
We obtain and study both analytically and numerically,
the equations defining  topologically stable
string-like objects. We check our results using two different approaches. Various
technical details are explicitly stated for the benefit of various readers.

\newpage
\section{Introduction}
Different problems in theoretical Physics suggest the existence of
{\it hidden sectors}. These consist of $SU(3)\times SU(2)\times U(1)$
singlet fields. Indeed, since many extensions of the Standard Model
propose the existence of additional product gauge groups (from Technicolor to
Heterotic string inspired models), we have
reasons to imagine that there are particles transforming under the
new gauge fields, but not under the familar local symmetries.

One possible way of coupling the 'hidden' and the 'visible'
sectors is mediated
by the Higgs field; these are called  Higgs-portals. These models propose
an interaction of the form $L_{int}\sim \alpha \Phi \Phi^* X X^*$,
where $\alpha$ is the coupling, $\Phi$
is the recently measured Higgs particle and $X$ a complex
field in the hidden sector
\cite{SZ},
\cite{PW}, \cite{DKM}.

On the other hand, different models of Supersymmetry (SUSY) breaking also
propose the existence of a hidden sector generating the dynamics
that breaks SUSY, hence avoiding the constraints imposed by sum-rules on
the masses of superpartners. In this case, the interest focuses on
the mechanism of communication between the hidden and visible sectors.
In the work \cite{DKM}, a mechanism was proposed that used a
renormalisable interaction between two $U(1)$ gauge fields, one visible
$A_\mu$ with curvature $F_{\mu\nu}$ and one hidden $C_\mu$, with
field strength $G_{\mu\nu}$. The interaction Lagrangian is
$L_{int}\sim \xi F_{\mu\nu} G^{\mu\nu}$ and was originally
proposed in \cite{Okun}-\cite{Holdom}.
This interaction
can be generated at arbitrarily high energies
and is not suppressed by powers of the scale,
due to the marginal character of the operator.
It was argued  by the authors of \cite{DKM} that this type of interactions
are generically induced by one-loop effects and  are
naturally expected to appear in String theory models.

As explained in \cite{Holdom}, gauge-gauge interactions of the form discussed
above,
can be diagonalised and the system becomes that of two decoupled gauge fields
at the expense of charging the matter fields (hidden and visible) under both
$U(1)'s$. This implies that the matter fields will develop a small charge---
if the parameter $\xi$ is small \footnote{If the number $\xi$ is not smaller
than $10^{-3}$ this interaction seems to be ruled out
experimentally \cite{Arkani}.}--- under the hidden and visible gauge symmetries.
Astrophysical observations place strong constraints on milli-charged matter
\cite{Davidson:2000hf}.
Hence, in what follows, we will consider that the hidden $U(1)$ group is
spontaneously broken. Indeed, if this hidden $U(1)$ were not broken,
'hidden photons' would interfere with and spoil
the process of nucleosynthesis.
Also, experiments such as laser polarisation or
light shining through a wall \cite{JR1} tightly constrain
the value of the parameter $\xi$.

The reader may wonder about more generic versions of the interaction,
involving for example non-Abelian gauge groups. It can be seen that
below the scale of breaking, the dynamics is well captured by the
Abelian interaction we are discussing, see \cite{Vachaspati}.
Hence, an accurate description of the low energy dynamics of our system consists
of two coupled Abelian Higgs Models. This is what we will do
in this paper to study the dynamics of topological objects.

Supersymmetric versions of the models with gauge-gauge interactions, like
the ones we are discussing above present various appealing features.
Among them,
that the hidden sector will provide (if R-parity preserving) a
candidate for dark matter. Hence, the dynamics of dark matter
and its interaction with the Standard Model particles
is another strong motivation to consider these models.

In this context, the problem of Physics that guided the present investigation
is the dynamics of 'hidden' (or 'dark') strings that appear in these systems.
These dark-strings \cite{Vachaspati}-\cite{Hyde}, have a tension of the
order of the symmetry breaking scale $\mu\sim (TeV)^2$. They are
not detectable
by their gravitational effects on the CMB,
but their coupling to the Standard Model fields (the
milli-charged particles already mentioned) can make the dark strings observable
via Bohm-Aharonov effect or by scattering of the Standard Model
particles with the string core \cite{Vachaspati}. See also \cite{Nelson:2011sf}
for various phenomenological aspects of these objects.
In this work, we will give a step towards the understanding of these
dark strings by investigating them as BPS objects.

To summarise our motivations and framework,
we have discussed the phenomenological interest
(either for Beyond the Standard Model scenarios or Dark Matter
models) of two type of interactions between hidden/dark and visible sectors;
gauge kinetic mixing term and Higgs portal interactions,
\begin{eqnarray}
 L_{GKM} = \frac\xi2 F_{\mu\nu}G^{\mu\nu},\;\;\;\;
L_{HP} = -\alpha \Phi^\dagger \Phi X^*X.
\label{gkm}
\end{eqnarray}
Now, with Supersymmetry, a gauge mixing term {\it automatically implies} the occurrence of
a Higgs portal.  Indeed, the  mixing of the
two auxiliary fields $D$, $D'$  belonging to the gauge multiplets
forces a mixing of the scalars in the chiral superfields.
We also need to add Fayet-Iliopoulos terms in such theories,
which orchestrate, via the Higgs mechanism, the needed spontaneous
breaking of
the gauge groups. A discussion of new Physics arising from this mixing
can be found in \cite{M2} and references therein.

As anticipated above, our system of interacting visible and hidden fields
will be accurately described at observable scales by two
Abelian Higgs Models coupled by a gauge kinetic mixing
interaction.
When these visible and/or  hidden $U(1)$ gauge
{symmetries are spontaneously broken,
the  classical equations of motion have vortex solutions,
which could be interpreted as visible and dark strings
\cite{Vachaspati}-\cite{Hyde}. Properties of such vortex
configurations arising in Abelian Higgs models with the gauge
kinetic mixing of eq.(\ref{gkm}) have been discussed
in  \cite{betti}-\cite{AS}. However, except for very
particular cases where the Ans\"atze for the gauge potentials
are equal \cite{betti}, no first order self-dual equations
have been found by establishing an energy
bound ``\`a la Bogomol'nyi \cite{Bogo} ---see the discussion in \cite{AS}.}

Here we propose a different approach to this problem.
As observed in \cite{dVS},  the   classical equations of motion
of the Abelian Higgs model  can be reduced to first-order
self-duality equations  when the gauge and
quartic scalar self-interaction  coupling constants obey
a relation naturally imposed by supersymmetry.
The theory then coincides exactly with a particular
bosonic sector of a highly supersymmetric parent theory.
The logic of this connection was understood after the work of
Olive and Witten on Bogomol'nyi equations for kinks and
dyons \cite{WO} and discussed in detail in the case of vortices
in \cite{ENS1}-\cite{ENS3}.

In more precise terms,  the gauge theory with spontaneous symmetry breaking
and a topological charge associated with  the vortex magnetic flux,
can be thought of as being part of
an ${\cal N}=2$ supersymmetric extension to the original model.
Its energy is bounded by the ${\cal N}=2$ central charge that is
proportional to the topological charge induced by the vortices
\cite{ENS1}. Furthermore, this bound is saturated when the
fields obey a set of first order equations, the
Bogomol'nyi-Prasad-Sommerfeld (BPS) equations.
Solutions of these equations naturally have finite energy and
therefore are well-suited to study objects like vortices.
These arguments, formulated for just one Abelian Higgs model,
extend naturally to a theory with a mixture of two such
models as described earlier, with gauge kinetic mixing and
Higgs portal type interactions.

{It is the purpose of this work to derive such extension.
The paper is organized as follows: in Section \ref{section2} we present the
${\cal N} = 2$ supersymmetric extension of a $(2+1)$ dimensional theory
in which two Abelian Higgs models are coupled through a gauge kinetic
mixing, showing how a Higgs portal interaction necessarily arises.
Starting from the supersymmetry transformations,
in Section \ref{section3} we obtain the Bogomol'nyi equations
and the energy of the system, bounded by the topological charge.
In Section \ref{section4} we present
the ${\cal N}=2$ supercharge algebra, which explains the connection
between the central charge and the visible and hidden magnetic fluxes. Both in sections
\ref{section3} and \ref{section4} various technical details will be made
explicit for the benefit of the phenomenologically-minded reader.
A careful numerical analysis of the vortex solutions is presented in
Section \ref{section5}. An
alternative derivation of the main results based
on diagonalization of the supersymmetric Action is described in Section
\ref{section6}.
A summary and discussion of our results is presented
in Section \ref{section7}. An Appendix complements the presentation.}

\section{The ${\cal N}=2$ supersymmetric model}\label{section2}

As discussed in the introduction,  vortex configurations in
a model with two $U(1)$ gauge fields $A_\mu$ and $C_\mu$, each one
coupled to complex scalar fields $s$ and $t$ respectively and
a gauge kinetic coupling were constructed in \cite{AS}.
Although the model is defined $d=3+1$ space-time dimensions,
vortex solutions naturally occur with rotational and traslational
invariance around a certain axis, thereby reducing the problem to $2+1$ dimensions,
where the supersymmetric extension can be easily formulated.
The Action governing the dynamics of the model discussed in \cite{AS}
can be  written in the form,
\begin{equation}
S=\int d^3x\left(  -\frac{1}{4}
F_{\mu\nu}F^{\mu\nu}-\half s^\dagger \hat{\Box} s-
V_1\left( s \right)  -\frac{1}{4}G_{\mu\nu}G^{\mu\nu}
-\half t^\dagger\tilde{\Box} t -V_2\left( t \right) +\frac{\xi}{2}F_{\mu\nu}G^{\mu\nu}\right),
\label{actione}
\end{equation}
where

\begin{equation}
\begin{split}
\hat{\Box}=\left( \partial_\mu-ieA_\mu\right)
\left( \partial^\mu -ieA^\mu\right)\text{, } \tilde{\Box}=
\left( \partial_\mu-igC_\mu\right) \left( \partial^\mu -igC^\mu\right)\text{, }\\
F_{\mu \nu}=\partial_\mu A_\nu - \partial_\nu A_\mu\text{, }
G_{\mu\nu}=\partial_\mu C_\nu -\partial_\nu C_\mu,
\end{split}
\end{equation}
and the potentials are,
\begin{align}
V_1\left(s \right)=\frac{a}{4}\left(\left|s \right| ^2-s_0^2 \right)^2\text{, }V_2\left(t \right)=\frac{b}{4}\left(\left| t\right| ^2-t_0^2 \right)^2.
\end{align}
The  second order equations of motion derived from this Action
in eq.\eqref{actione} reveal a very rich structure of vortex
solutions and the dependence on the parameters of the theory
shows a large variety of phenomena that could be of
interest in connection to the problems described
in the introduction. Solutions have  been found numerically,
by solving the second order equations of motion,
since it was not obvious how to find Bogomol'nyi equations for this system,
due to the gauge-mixing term.

As it is well known the existence of Bogomol'nyi
equations is closely related to the existence of a
{ (${\cal N} = 2$) extensions of purely bosonic
models exhibiting  kinks, vortices or monopole solutions
\cite{dVS}-\cite{ENS3}. In general, such
extensions are only possible for very particular symmetry breaking
potentials (or even  for vanishing  potential, as it is the case
for models with monopole  or dyon solutions). Indeed, the form of these potentials
is a requirement
for supersymmetry to hold}.
Then, a practical way to find models accepting Bogomol'nyi equations
is to formulate the problem in superspace, without an explicit potential,
and let the supersymmetry algebra guide us towards the correct form for the Action,
that will  have its coupling constants in the right ratios.

\subsection{The supersymmetric action}
{Three-dimensional $N=2$ supersymmetry can be obtained
by dimensional reduction from the equivalent $N=1$ four-dimensional theory}.
In practice, the supersymmetry representations are tractable by
themselves, though we will keep the connection with
the higher dimensional theory in mind during the process.

Let us first set up the technical tools needed to construct the SUSY extension
of the model in eq.(\ref{actione}).
In order to enforce two copies of supersymmetry
one needs two sets of Grassmann variables
$\theta^\alpha$, $\bar{\theta}^\alpha$. Note that there is only one spinor representation
in three-dimensions.
We work with the $(+--)$ signature, contract indices
using $\epsilon_{\alpha\beta}$, and choose  the following $\gamma$-matrices,
\begin{equation}
\gamma^0=\left( \begin{array}{cc}
1 & 0  \\
0 & -1
\end{array} \right)\text{, }\gamma^1=\left( \begin{array}{cc}
0 & 1  \\
-1 & 0
\end{array} \right)\text{, }\gamma^2=\left( \begin{array}{cc}
0 & i  \\
i & 0
\end{array} \right),
\end{equation}
so that $\gamma^\mu \gamma^\nu=\eta^{\mu\nu}+i\epsilon^{\mu\nu\rho}\gamma_\rho$.
We will abuse the bar notation: over spinorial coordinates, this refers to
two independent quantities, but if $\lambda$ is a (complexified)
fermion field, then $\bar{\lambda}=\gamma^0\lambda^\dagger$ is not independent.
We write the super-derivatives analogously to those in the $d=3+1$ dimensional case
\begin{equation}
\mathcal{D}_\alpha=\partial_\alpha+i\left( \gamma^\mu\bar{\theta}\right)_\alpha\partial_\mu\text{, }\bar{\mathcal{D}}_\alpha=\bar{\partial}_\alpha +i\left( \theta\gamma^\mu\right)_\alpha\partial_\mu\text{, }\lbrace\mathcal{D}_\alpha,\bar{\mathcal{D}}_\beta \rbrace=2i\gamma^\mu_{\alpha\beta}\partial_\mu.
\end{equation}

We also use the standard chiral hypermuliplet representation $\Phi$ for the matter fields.
It contains one complex scalar $s$ and one full Dirac spinor i.e.
two independent Majorana spinors written as one complex spinor $\psi$,
along with a complex auxiliary F-term. This auxiliary field
will not play any relevant role, since as we shall see, our bosonic theory is realised
without a superpotential. We define these individual components by
the action of the super-derivatives on the field,
evaluated at $\theta=\bar{\theta}=0$; we denote the evaluation with $\rvert$. Hence,
\begin{align}
\Phi\rvert=s\text{, }\mathcal{D}_\alpha\Phi\rvert= \bar{\psi}_\alpha \text{, }\mathcal{D}_\alpha\mathcal{D}_\beta\Phi\rvert=\gamma^\mu_{\alpha\beta}\partial_\mu s+\epsilon_{\alpha\beta}F\text{, } \mathcal{D}_\alpha\mathcal{D}^\beta\bar{\mathcal{D}}_\beta \Phi \rvert=(\slashed{\partial}\bar{\psi})^\dagger_\alpha
\end{align}
and
\begin{align}
\Phi^\dagger\rvert=s^\dagger\text{, }\bar{\mathcal{D}}_\alpha\Phi^\dagger\rvert= \psi_\alpha \text{, }\bar{\mathcal{D}}_\alpha\bar{\mathcal{D}}_\beta\Phi^\dagger\rvert=(\gamma^\mu_{\alpha\beta}\partial_\mu s+\epsilon_{\alpha\beta}F)^\dagger\text{, } \mathcal{D}_\alpha\bar{\mathcal{D}}^\alpha\bar{\mathcal{D}}_\beta \Phi \rvert=(\slashed{\partial}\psi)_\alpha.
\end{align}
The other matter superfield $\Psi$ is treated identically  and contains the scalars $t$,
a Dirac fermion $\sigma$ and an auxiliary field $G$.

The vector multiplet $U$ possesses one real scalar $M$,
two Majorana fermions written as one complex fermion $\lambda=\chi+i\rho$
and one gauge field $A_\mu$, along with several auxiliary fields, all but
one of which we can choose to ignore in the
Wess-Zumino gauge.
The only auxiliary field we cannot gauge away
in this multiplet, we will call it $D$. Thus we define,
\begin{equation}
\mathcal{D}_\alpha\bar{\mathcal{D}}_\beta U\rvert= \epsilon_{\alpha\beta}M+\gamma^\mu_{\alpha\beta}A_\mu\;\text{, }~~~~ \mathcal{D}^2\bar{\mathcal{D}}^2U\rvert=D\;\text{, }~~~~ \mathcal{D}^2\bar{\mathcal{D}}_\alpha U\rvert=\bar{\chi}_\alpha\;\text{, }~~~~ \mathcal{D}_\alpha\bar{\mathcal{D}}^2U\rvert=\rho_\alpha.
\end{equation}

The standard gauge-invariant curvature tensor $W_\alpha$ and its conjugate can be
defined to generate the canonical kinetic term.  In three dimension there
exists an extra representation called the linear
multiplet $\Sigma= \bar{\mathcal{D}}\mathcal{D}U$ which is real
and obeys $\mathcal{D}^2\Sigma=\bar{\mathcal{D}}^2\Sigma=0$.
This field proves to be more convenient for component definitions
as it contains all the degrees of freedom at once.
Thus, using our previous results we define,
\begin{equation}
\Sigma\rvert=M\;\text{, }~~~~\mathcal{D}_\alpha \bar{\mathcal{D}}_\beta
\Sigma\rvert=\epsilon_{\alpha\beta}D+i\gamma^\mu_{\alpha\beta}\left( i\epsilon_{\mu\nu\rho}\partial^\nu A^\rho+\partial_\mu M\right)\;\text{, }
\end{equation}
\begin{equation}
\mathcal{D}_\alpha\Sigma\rvert=\bar{\chi}_\alpha\;\text{, }~~~~
\bar{\mathcal{D}}_\alpha\Sigma\rvert=\rho_\alpha\text{, }~~~~ \mathcal{D}_\alpha\bar{\mathcal{D}}^2\Sigma\rvert=i\gamma^\mu_{\alpha\beta} \partial_\mu \chi_\beta\;\text{, }~~~~
\mathcal{D}^2\bar{\mathcal{D}}_\alpha\Sigma\rvert=i\gamma^\mu_{\beta \alpha}\partial_\mu \bar{\rho}_\beta.
\end{equation}

A second multiplet to describe the 'hidden' sector, the analogous to $\Sigma$,
will be called $\Upsilon$   with bosonic fields
$C_\mu$, $N$, auxiliary field $d$ and fermions $\tau=\zeta+i\omega$.
With these definitions, we can now write the superspace Action for our model;

\begin{equation}
S_{\mathcal{N}=2}=\int d^3x d^2\theta d^2\bar{\theta} \left( \frac{1}{4}\Sigma\Sigma+\frac{1}{4}\Phi^\dagger e^{-ieU}\Phi
+ \frac{1}{4}\Upsilon\Upsilon+\frac{1}{4}\Psi^\dagger e^{-igV}\Psi -\frac{\xi}{2}\Sigma\Upsilon +\frac{ies^2_0}{2}U +\frac{igt^2_0}{2}V \right)
\label{cici}.
\end{equation}
The final two terms in the action are Fayet-Iliopoulos terms introduced to
achieve the phenomenologically required
spontaneous gauge symmetry breaking.
Noting that  for any field $\Delta$
\be
\int d^2\theta d^2\bar{\theta} \Delta \;\hat{=} \; \mathcal{D}^2\bar{\mathcal{D}}^2\Delta\rvert,
 \ee
it is then straightforward to use the previous relations,
expand the superfields in components and
obtain the SUSY-completed action for the pair of
coupled Abelian Higgs system.
We assume that the spinors, which all come in pairs,
have been complexified according to $\lambda=\chi+i\rho$
and $\tau=\zeta+i\omega$.
It is convenient to write the complete action in the form,
\begin{equation}
S_{{\cal N}=2}=S_1\left(A,s,M,\psi,\lambda \right)+
S_2\left(C,t,N,\sigma, \tau \right)-\xi S_{int}\left(A,M, C,N,\lambda,\tau, d, D \right),
\label{action}
\end{equation}
with
\begin{eqnarray}
S_1 &=&\int d^3x \left(-\frac{1}{4}F_{\mu\nu}F^{\mu\nu}-\half s^\dagger \hat{\Box}
s+\frac{1}{2}D^2-\frac{ie}{2}\left(|s|^2-s_0^2\right)D \right.\nonumber\\
&&\left. -\frac{1}{2}M\Box M-\frac{1}{4}M^2|s|^2+\frac{i}{2}\bar{\psi}\slashed{D}_A\psi
+\frac{i}{2}\bar{\lambda}\slashed{\partial}\lambda-\frac{1}{2}M\bar{\psi}\psi
-\frac{e}{2}\left(\bar{\psi}\lambda s+s^\dagger\bar{\lambda}\psi\right)\right).
\label{s1zzz}
\end{eqnarray}
{The Action $S_1$}, describing the 'visible' sector, is invariant under the following
transformations with infinitesimal anticommuting  complex parameters $\eta$,
\[
\delta s= \bar{\eta}\psi\text{ , }\delta\psi=-i\gamma_\mu\eta D^\mu_A s-\eta M s\text{ , }\delta M= \bar{\eta}\lambda-\bar{\lambda}\eta\text{ , }\delta A_\mu=-i\bar{\eta}\gamma_\mu \lambda+i\bar{\lambda}\gamma_\mu \eta \]
\begin{equation}\delta D=\partial^\mu\left(\eta\gamma_\mu\bar{\lambda}-\bar{\eta}\gamma_\mu\lambda \right) \text{ , }
\delta \lambda=-\epsilon^{\mu\nu\rho}\partial_\mu A_\nu \gamma_\rho\eta-i\slashed{\partial}M\eta-iD\eta.
\label{susy1zz}
\end{equation}
%

The second ('dark') sector has a similar action,
\begin{eqnarray}
S_2 &=& \int d^3x \left( -\frac{1}{4}G_{\mu\nu}G^{\mu\nu}-\half t^\dagger \tilde{\Box} t+\frac{1}{2}d^2-\frac{ig}{2}\left(|t|^2-t_0^2\right)d
\right. \nonumber\\
&&- \left.\frac{1}{2}N\Box N-\frac{1}{4}N^2|t|^2
+\frac{i}{2}\bar{\sigma}\slashed{D}_C\sigma+\frac{i}{2}\bar{\tau}\slashed{\partial}\tau-\frac{1}{2}N\bar{\sigma}\sigma-
\frac{g}{2}\left(\bar{\sigma}\tau t+t^\dagger\bar{\tau}\sigma\right) \right).
\label{s2zzz}\end{eqnarray}
The Action {$S_2$} is invariant under the following transformations,
\[
\delta t= \bar{\eta}\sigma\text{ , }\delta\sigma=-i\gamma_\mu\eta D^\mu_C t-\eta N t\text{ , }\delta N= \bar{\eta}\tau-\bar{\tau}\eta\text{ , }\delta C_\mu=-i\bar{\eta}\gamma_\mu \tau +i\bar{\tau}\gamma_\mu \eta \]
\begin{equation}
\delta d=\partial^\mu\left(\eta\gamma_\mu\bar{\tau}-\bar{\eta}\gamma_\mu\tau \right) \text{ , }
\delta \tau=-\epsilon^{\mu\nu\rho}\partial_\mu C_\nu \gamma_\rho\eta-i\slashed{\partial}N\eta-id\eta.
\label{susy2zz}\end{equation}

The term coupling the two visible and dark sectors,
which is invariant under both sets of transformations reads
\begin{equation}
\begin{split}
S_{int}=\int d^3x\left( -\frac{1}{2}F_{\mu\nu}G^{\mu\nu} -\frac{1}{2}\left( M\Box N +N\Box M\right)+\frac{i}{2}\left(\bar{\lambda}\slashed{\partial}\tau+\bar{\tau}\slashed{\partial}\lambda\right)+ dD\right).
\end{split}
\label{communicating}
\end{equation}


Let us discuss the form of the scalar potential derived from the auxiliary fields.

\subsection{The scalar potential}\label{scalapotential}
In order to obtain the symmetry breaking potential we have to solve  the equations
of motion for auxiliary fields $D$ and $d$ whose
contribution will be collected in $L_{Dd}$,
\begin{equation}
L_{Dd} =
\frac{1}{2}D^2+\frac{1}{2}d^2-\xi dD-\frac{ie}{2}\left(|s|^2-s_0^2\right)D-\frac{ig}{2}\left(|t|^2-t_0^2\right)d,
\end{equation}
or
\begin{equation}
L_{dD}= \frac{1}{2}\left( \begin{array}{cc}
D & d
\end{array} \right) \left(\begin{array}{cc}
1 & -\xi \\
-\xi & 1
\end{array}  \right) \left( \begin{array}{c}
D \\
d
\end{array} \right) - \left( \begin{array}{cc}
D & d
\end{array} \right) \left( \begin{array}{c}
\frac{ie}{2}\left(|s|^2-s_0^2\right) \\
\frac{ig}{2}\left(|t|^2-t_0^2\right)
\end{array}
 \right).
 \label{entra}
\end{equation}
The extrema of this quadratic system is given by,
\begin{equation}
\left( \begin{array}{c}
D \\
d
\end{array} \right)= \frac{1}{1-\xi^2}\left(\begin{array}{cc}
1 & \xi \\
\xi & 1
\end{array}  \right)
\left( \begin{array}{c}
\frac{ie}{2}\left(|s|^2-s_0^2\right)  \\
\frac{ig}{2}\left(|t|^2-t_0^2\right)
\end{array} \right)
\label{DdDd},
\end{equation}
which gives,
\bea
D &=&  \frac{i}2 \frac1{1 - \xi^2}\left( e(|s|^2-s_0^2) + \xi g (|t|^2-t_0^2) \right) \label{D1}\\
d &=&  \frac{i}2 \frac1{1 - \xi^2}\left( e\xi (|s|^2-s_0^2) +   g (|t|^2-t_0^2) \right) \label{D2}.
\eea
Substituting this expression in eq.\eqref{entra} one gets,
\begin{equation}
L_{Dd} = -\frac{1}{2\left(1-\xi^2 \right) }\left( \begin{array}{cc}
\frac{ie}{2}\left(|s|^2-s_0^2\right) &
\frac{ig}{2}\left(|t|^2-t_0^2\right)
\end{array} \right)
\left(\begin{array}{cc}
1 & \xi \\
 \xi & 1
\end{array}  \right)
\left( \begin{array}{c}
\frac{ie}{2}\left(|s|^2-s_0^2\right) \\
\frac{ig}{2}\left(|t|^2-t_0^2\right)
\end{array} \right).
\end{equation}
So that finally the scalar potential takes the form
\begin{equation}
V[s,t] = \frac{1}{2\left(1-\xi^2 \right) } \left( \frac{e^2}{4}\left(|s|^2-s_0^2\right)^2 +\frac{g^2}{4}\left(|t|^2-t_0^2\right)^2
+\frac{eg\xi}{2}\left(|s|^2-s_0^2\right)\left(|t|^2-t_0^2\right)
\right).
\label{cahpotential}
\end{equation}
Clearly this reduces to the usual, decoupled form in the
case $\xi=0$. {To avoid singularities, the
parameter $\xi$ should be constrained so that  $|\xi|<1$. In fact, as discussed
in \cite{AS}, the existence of solutions with the appropriate asymptotic
boundary conditions imposes such constraint.}


The equations for the  extrema of the potential read
\begin{equation}
 \begin{array}{c}
e^2|s|\left(|s|^2-s_0^2 \right)+ge\xi|s|\left(|t|^2-t_0^2\right)=0\\
~ \nonumber
\\
g^2|t|\left(|t|^2-t_0^2 \right)+eg\xi |t|\left(|s|^2-s_0^2 \right) =0
\end{array}
\end{equation}
and the Hessian matrix is given by
\begin{equation}
{\cal H} =
 \frac1{2(1 - \xi^2)}\left( \begin{array}{cc}
3e^2 |s|^2+e^2 s_0^2+eg\xi \left( |t|^2 -t_0^2 \right)   &2eg\xi |s||t| \\
2eg\xi |s||t| & 3g^2|t|^2+g^2t_0^2+eg\xi \left( |s|^2 -s_0^2 \right)
\end{array} \right)
\end{equation}
There are 4 particular types of critical points of interest:
\begin{itemize}
\item $|s|=|t|=0$: Maximum, the vacuum is unstable.
\item $|s|=0$, $|t|=\sqrt{t_0^2+\frac{e\xi}{g}s_0^2} $: Saddle point
\item $|s|=\sqrt{s_0^2+\frac{g\xi}{e}t_0^2}$, $|t|=0$: Saddle point\footnote{These saddle points can disappear if $\xi$ is negative, and if the VEVs and coupling constants take such values as to make the inside of these square roots negative.}
\item $|s|=s_0$, $|t|=t_0$: This is the true minimum.
\end{itemize}
This verifies that we have perturbed each of the
Abelian Higgs Models. Let us now turn to the main point of this procedure, the BPS equations.

\section{Bogomol'nyi equations}\label{section3}
In this section we will derive BPS equations for the system described in eq.(\ref{actione}).
We will follow the well-known procedure, basically imposing that a purely bosonic
configuration preserves part of the SUSY of the Action in eq.(\ref{action}).
\subsection{BPS states and equations}
Starting from the ${\cal N} = 2$ supersymmetric Action in eq. \eqref{action}
we are interested in finding a purely bosonic action in which
gauge fields and Higgs scalars in the visible and hidden sectors
are coupled in such a way that first order BPS equations do exist. This will be achieved by
 enforcing that only the bosonic part of the ${\cal N} = 2$ supersymmetric Action
eq. \eqref{action} subsists. Following this procedure one ends
with extra adjoint (i.e. ungauged) scalars for each gauge group, which we also require to be zero.

The rationale behind this procedure
is well-known: we are free to impose that physical states have only
the degrees of freedom we desire (gauge particles and squarks)
and not break supersymmetry completely by imposing that the supersymmetric
variations of each vanishing field is identically zero.
Such states are called BPS states, because they saturate the
Bogomol'nyi lower bound for the total energy of the system.
Recalling the supersymmetric variations previously established,
and imposing that only our physical fields appear,
we obtain the following set of equations, for an arbitrary infinitesimal spinor $\eta$
by demanding that the variations of the fermion fields vanish;

\begin{equation}
\begin{split}
-i\gamma_\mu\eta D^\mu_A s=0\text{ , }-\epsilon^{\mu\nu\rho}\partial_\mu A_\nu \gamma_\rho\eta-iD\eta=0,\\
-i\gamma_\mu\eta D^\mu_C t=0\text{ , }-\epsilon^{\mu\nu\rho}\partial_\mu C_\nu \gamma_\rho\eta-id\eta=0.
\end{split}
\label{unos}
\end{equation}

Furthermore, to write the equations {leading to magnetic vortex solutions,
we shall
impose time-independence of our solutions and use the gauge choice $A_0=C_0=0$.
Rewriting the above we get,}
\begin{equation}
\begin{split}
\left( -i\gamma_1\eta D^1_A-i\gamma_2\eta D^2_A\right)  s=0\text{ , }-\epsilon^{0ij}\partial_i A_j \gamma_0\eta-iD\eta=0\\
\left( -i\gamma_1\eta D^1_C-i\gamma_2\eta D^2_C\right)  t=0\text{ , }-\epsilon^{0ij}\partial_i C_j \gamma_0\eta-id\eta=0
\end{split} \label{dos}
\end{equation}
{Then, we multiply the scalar equations by $\gamma^1$
and notice that in all cases
the equations are proportional to the identity or $\gamma^0$ times the arbitrary spinor $\eta$.
 The resulting  equations, acting on each of the spinor components,
differ only by a sign change, so clearly they cannot be satisfied both
at the same time but are valid for a definite sign choice
in all cases. This is why precisely half of supersymmetry is broken
(either the symmetry associated with
$\eta_+$ or $\eta_-$ must be selected).
After using the eqs.\eqref{D1}-\eqref{D2} for $D$ and $d$, the Bogomol'nyi equations read,
}

\begin{equation}
\begin{split}
i\epsilon_{ij}D^i_A s=\pm \left( D^A_j s\right)^\ast
\text{ ,  }{\epsilon_{ij}\partial_i A_j} =\pm{\frac12}\frac{1}{1-\xi^2}\lbrace e\left(|s|^2-s_0^2\right)+g\xi \left(|t|^2-t_0^2 \right)\rbrace \\
i\epsilon_{ij}D^i_C t=\pm\left( D^C_j t\right)^\ast
\text{ , }{\epsilon_{ij}\partial_i C_j}=\pm {\frac12}\frac{1}{1-\xi^2}\lbrace g\left(|t|^2-t_0^2\right)+e\xi \left(|s|^2-s_0^2 \right)\rbrace
\end{split}
\label{tress}
\end{equation}
\subsection{The BPS bound for the energy}
As a consitency check, we shall re-derive the self-dual equations \eqref{tress}
following the Bogomol'nyi approach \cite{Bogo}.
This consist in writing the energy as a manifestly positive quantity
plus a topological term. In this way a bound for the energy can be obtained.
A set of first order equations precisely saturate this bound.

Once the extra scalars $M$ and $N$, as well as the fermion fields,
are put to zero, and the auxiliary fields $D$ and $d$ are written in
terms of the dynamical bosonic fields, the Lagrangian associated to the Action in
eq. \eqref{action} reads,

\bb
\mathcal L= -\frac{1}4 F_{\mu\nu} F^{\mu\nu}-\frac{1}4G_{\mu\nu} G^{\mu\nu}+ \frac{\xi}2F_{\mu\nu} G^{\mu\nu}+ \frac12\mathcal |D_\mu (A)s|^2+\frac12\mathcal |D_\mu (C)t|^2- V.
\label{lag}
\ee
Where
\bb
V= \frac{1}{2(1-\xi^2)} \left(\frac{e^2}4\left(|s|^2-s_0^2\right)^2 +\frac{g^2}4\left(|t|^2-t_0^2\right) ^2 +\frac{eg\xi}2 \left(|s|^2-s_0^2\right) \left(|t|^2-t_0^2\right)\right).
\ee

We are looking for static vortex-like classical solutions to
the equation of motion in the gauge $A_0 = C_0 = 0$, which have
quantized magnetic fluxes  associated to $A_i$ and $C_i$
\be
\Phi_A = \oint A_i dx^i = \frac{2\pi n}e \;, \;\;\;\; \Phi_C = \oint C_i dx^i = \frac{2\pi k}g \;, \;\;\;   n,k    \in \mathbf{Z}.
\label{xxy}\ee
To this end, it is convenient to  introduce dimensionless variables,
\bb
x_i\rightarrow x_i/e s_0, \,\,\,\,\,\,\, A_i\rightarrow A_i s_0,
\,\,\,\,\,\,\,  s \rightarrow s s_0, \,\,\,\,\,\,\, C_i\rightarrow C_i s_0,
\,\,\,\,\,\,\, t\rightarrow t s_0
\label{red1}
\ee
and the coupling constants  and gauge field masses ratios,
\be
e_r\equiv g/e\;, \;\;\;\;  \mu^2 \equiv e_r^2 t_0^2/s_0^2.
\label{red2}
\ee
In terms of these fields, the total energy of the system is
\bb
\frac{E}\ell=s_0^2 \int d^2 x\left\{\frac{B_A^2}2+\frac{B_C^2}2  +\half |\partial_is-iA_is|^2+\half|\partial_it-ie_rC_i t|^2-\xi B_AB_C+V(|s|)+V(|t|)+V_{\rm int}\right\}.
\label{ip}
\ee
with
 $\ell = 1/(e s_0)$ and magnetic fields $B_A$ and $B_C$ defined as,
\be
 B_A =   \epsilon_{ij}\partial_{i}A_j = F_{12} \; ,  \;\;\;
 B_C  = \epsilon_{ij}\partial_{i}C_j = G_{12}.
 \ee
 Concerning the potentials, we have
\bb
V(|s|)=\frac{1}{8(1-\xi^2)}\left(|s|^2-1\right)^2,\,\,\,\, V(|t|)=\frac{e_r^2}{8(1-\xi^2)}\left(|t|^2-\left(\frac{\mu}{e_r}\right)^2\right)^2,
\label{V1}
\ee
and
\bb
V_{\rm int}=\frac{e_r\xi}{4(1-\xi^2)}\left(|s|^2-1\right)\left(|t|^2-\left(\frac{\mu}{e_r}\right)^2\right).
\label{V2}
\ee
{To study the  equations of motion, let us introduce a cylindrically
symmetric ansatz for the fields, in terms of radial functions as}
\bb
A_\varphi= \frac{\alpha(r)}r, \,\,\,\,\,\, C_\varphi= \frac{\gamma(r)}{e_r r},\,\,\,\,\,\,  s= \rho(r) e^{i\varphi},\,\,\,\,\,\, t= p(r) e^{i\varphi}.
\ee
The energy density in terms of the radial functions takes the form

{\bea
\mathcal E &=& \frac{1}{2r^2}\left( \frac{d \alpha}{dr} \right)^2+ \frac{1}{2r^2 e_r^2}\left( \frac{d \gamma }{dr}\right)^2+ \frac12\left(\left( \frac{d \rho}{dr} \right)^2
+ (\alpha - 1)^2\frac{\rho^2}{r^2}
\right) \nonumber\\
&& +\frac12\left(\left( \frac{d p}{dr} \right)^2 + (\gamma - 1)^2\frac{p^2}{r^2} \right)
-\frac{\xi}{e_r}  \frac{d \alpha}{dr}  \frac{d \gamma}{dr} +V(\rho)+V(p)+V_{\rm int}
\label{ener}
\eea}
{It is clear from the equation above
that if the parameter $\xi/e_r$ gets bigger, the magnetic energy
diminishes.
}

{Concerning the second order radial equation of motion they read,}
\eqb
&& {r \frac{d}{dr} \left[\frac{1}r \frac{d}{dr} \left(\alpha + \frac{\xi}{e_r} \gamma \right) \right]+  (1-\alpha) \rho^2=0, }\label{color}\\
&& {r \frac{d}{dr} \left[\frac{1}r \frac{d}{dr} \left(\gamma +{\xi}{e_r} \alpha \right)
\right]+  e_r^2(1-\gamma) p^2=0,}\\
&& \frac{1}r \frac{d}{dr} \left[r \frac{d}{dr}\rho \right]-{\frac{1}{r^2}}
\left(\alpha -1\right)^2 \rho- \frac{1}{2\left(1-\xi^2\right)}
\left(\rho^2-1+e_r\xi\left(p^2-\frac{\mu^2}{e_r^2} \right)\right)\rho=0,\\
&& \frac{1}r \frac{d}{dr} \left[r \frac{d}{dr}p \right]-{\frac{1}{r^2}}
\left(\gamma -1\right)^2 p- \frac{e_r^2}{2\left(1-\xi^2\right)}
\left(p^2-\frac{\mu^2}{e_r^2}+\frac{\xi}{e_r}\left(\rho^2-1 \right)\right)p=0.
\label{colores}
\eqf
Notice that in the limit case
of $\xi=e_r=\mu=1 $, the kinetic term for the combination of gauge fields
$\alpha-\gamma$ decouples.

In view of the  well-known connection
between BPS states and the Bogomol'nyi
bound to the energy \cite{WO}-\cite{ENS1},
we know that the solution to  eqs.\eqref{tress}
also solve the equations of motion in eqs. \eqref{color}-\eqref{colores}.
Indeed, the energy in eq. \eqref{ip} can be rewritten as a manifestly positive quantity,
by "completing the square", which in this case is not just
a square but a positive definite quadratic form.
\begin{eqnarray}
 E &= &\int d^2x  \left(\frac12 \left(B_A \mp{\frac12}\frac1{(1 - \xi^2)}\left(\!e(|s|^2 - s_0^2)+\xi g(|t|^2 - t_0^2)  \right)\right)^2 \right.
   \nonumber\\
   && +  \frac12 \left( B_C \mp {\frac12}\frac1{(1 - \xi^2)} \left( g(|t|^2 - t_0^2) + \xi e(|s|^2 - s_0^2)\right)\!
\right) ^2  \nonumber\\
 &&-  \xi\left(B_A \mp {\frac12}\frac1{(1 - \xi^2)} \left( e(|s|^2 - s_0^2) + \xi  g(|t|^2 - t_0^2)\right)  \right)
 \left(B_C \mp {\frac12}\frac1{(1 - \xi^2)}\left(  g(|t|^2 - t_0^2) + \xi e(|s|^2 - s_0^2)\right) \right)  \nonumber\\
&&+ \left.\left( \frac12 \left\vert\vphantom{\int}\varepsilon_{ij}D_j[A]s_a \mp \epsilon_{ab}D_i[A] s_b\right\vert^2+\frac12\left|\vphantom{\int}\varepsilon_{ij}D_j[C]t_a \mp \epsilon_{ab}D_i[C] t_b\right|^2   \pm \partial^i {\cal J}_i  \vphantom{\frac1{x^2}}\right) \right) .
\label{over}
\end{eqnarray}
Here we have written the real and imaginary components of the scalar fields as $ s_a$ with $a=1,2$ respectively and defined covariant derivatives   as
\be
D_is_a = \partial_i s_a + \epsilon_{ab} A_i s_b  \;, \;\;\;
\ee
and analogously for the hidden scalar field $t_a$.
All but the last term in this energy are part of a positive-definite quadratic form,
so that this part of the energy is always positive.
Concerning the last term ---generated to complete the form---
the current ${\cal J}_i$  is given by
\bea
{\cal J}_i &=&   \varepsilon_{ij}\left(\varepsilon_{ab}\left(s_a D_j[A] s_b + A_j\right)\right)  + \varepsilon_{ij}(\varepsilon_{ab}(t_a D_j[C] t_b + \frac1{e_r}C_j)).
\eea
 Using Stoke's theorem and using
the fact that covariant derivatives vanish at
infinity (since we impose finite energy for our solutions)
one ends with the Bogomol'nyi bound for the energy which written in terms of the original  (unscaled) fields reads
 \be
 E  \geq e s^2_0|\Phi_A| + e t_0^2|\Phi_C| = s_0^2 {2\pi |n|}+
 t_0^2  {2\pi |k|}
 \label{boundit}
 \ee
The bound is attained when   each one of the squares in
eq.\eqref{over} vanish, this leading precisely to the already obtained
equations \eqref{tress}, that
written in terms of the Ansatz {read}
{
\eqb
\frac{d\rho}{dr}&=&\left(-1+\alpha\right)\frac{\rho}r,\label{351}\\
\frac{dp}{dr}&=&\left(-1+\gamma \right) \frac{p}r,\label{352}\\
\frac{1}r \frac{d\alpha}{dr}&=&\pm \frac{1}{2(1-\xi^2)} \left[ (\rho^2-1) + e_r \xi (p^2-\frac{\mu^2}{e_r^2})\right],\\
\frac{1}r \frac{d\gamma}{dr}&=&\pm \frac{1}{2(1-\xi^2)} \left[ e_r (p^2-\frac{\mu^2}{e_r^2}) + \xi (\rho^2-1^2) \right].\label{bps353}
\eqf
}

Note that if $\xi=\pm 1$ in eq. (\ref{over}) the purely positive part
of the energy degenerates and can be factorised again into
a simpler expression.

In summary, we have obtained the BPS equations for our system in eq.(\ref{actione}).
We will now study how its topological charge can be re-obtained using a different approach,
based on the SUSY algebra.

\section{Supercharges}\label{section4}

The ${\cal N}=2$ action of eq. \eqref{action}  is invariant under
two super-transformations, detailed in eqs.(\ref{susy1zz}) and (\ref{susy2zz}).
The fermionic Noether charge associated with these invariances is given by,
\begin{equation}
\bar{Q}\eta =\sum_{\zeta\in\text{fermions}}\dfrac{\delta S}{\delta\left( \partial_0\zeta\right) }\delta_\eta\zeta .
\end{equation}
This gives the following expressions
\begin{eqnarray}
\bar{Q} &=& \int d^2x \left(\left( \bar{\lambda}-\xi\bar{\tau}\right) \gamma^0 \left( -\frac{1}{2} \epsilon^{\mu\nu\rho}F_{\mu\nu} \gamma_\rho-i\slashed{\partial}M-iD \right)
+\left( \bar{\tau}-\xi\bar{\lambda}\right) \gamma^0 \left( -\frac{1}{2} \epsilon^{\mu\nu\rho}G_{\mu\nu} \gamma_\rho-i\slashed{\partial}N-id) \right)\right. ,
\nonumber\\
&& \left. +\,\bar{\sigma}\gamma^0\left( -i\left( \slashed{\partial}-ig\slashed{C}\right)t-\frac{1}{2}Nt \right)  
 +\bar{\psi}\gamma^0\left( -i\left( \slashed{\partial}-ie\slashed{A}\right)s-\frac{1}{2}Ms \right)
 \vphantom{\frac{a^3}{2}}  \right).
 \label{sucha2}
 \\
 ~\nonumber\\
Q &=& \int d^2x \left( \left( -\frac{1}{2} \epsilon^{\mu\nu\rho}\left( F_{\mu\nu}-\xi G_{\mu\nu}\right)  \gamma_\rho+i\slashed{\partial}(M-\xi N)+i(D-\xi d)^\ast \right)\gamma^0 \lambda\right. \nonumber
\\
&& +\left( -\frac{1}{2} \epsilon^{\mu\nu\rho}(G_{\mu\nu}-\xi F_{\mu\nu}) \gamma_\rho+i\slashed{\partial}(N-\xi M)+i(d-\xi D)^\ast \right)\gamma^0 \tau
\nonumber\\
&&\left. +\left( i\left( \slashed{\partial}+ig\slashed{C}\right)t^{\ast}-\frac{g}{2}Nt^\ast \right)\gamma^0 \sigma +\left( i\left( \slashed{\partial}+ie\slashed{A}\right)s^{\ast}-\frac{e}{2}Ms^\ast \right)\gamma^0 \psi \vphantom{\frac{a^3}{2}}  \right)
 \label{sucha1}
\end{eqnarray}

These charges have been rewritten in terms of the fermionic fields (in $Q$) and their conjugate momenta (in $\bar{Q}$) so as to more easily impose canonical anti-commutation relations later on. We have not yet substituted for the on-shell value of $D$ and $d$ given in eq.\eqref{DdDd}.

We can now rederive
  the Bogomol'nyi bound on the energy of the system, and its saturatation by self-dual equations from the supercharge algebra.
{Indeed, as it is well known \cite{dVS}-\cite{ENS3}} that
in the supersymmetry context the Bogomol'nyi equations imply
that the total energy of the system is bounded below by the central
charge of the theory which {is proportional} to the topological charge associated to the solutions of the self-dual equations     (in our case the number of vortex flux units  of both sectors).

The Supersymmetry algebra dictates that, in the rest frame
\begin{equation}
\left\lbrace Q_\alpha,\bar{Q}_\beta\right\rbrace = \gamma^0_{\alpha\beta}E+\mathbbm{1}_{\alpha\beta}T,
\end{equation}
here $E$ is the total energy of the system and $T$ is the central charge. Squaring and tracing over this equation it is easy to get,
\beq
E\geq|T|,
\label{desig}
\eeq
which is the promised bound. Now, using the explicit form
of the supercharges in eqs. (\ref{sucha2})-(\ref{sucha1}),
 we can calculate explicitly  $E$ and $T$
Since we will eventually only keep bosonic terms,
we need to impose that the fermions and their canonical conjugate obey usual
Hamiltonian mechanics commutation relations.
Because of the presence of the mixing term,
the conjugate momentum for the gauginos is not trivial. Indeed, we find that
\begin{equation}
\dfrac{\delta S}{\delta\partial_0\lambda}=\left( \bar{\lambda}-\xi\bar{\tau}\right)\gamma^0 \text{ , }\dfrac{\delta S}{\delta\partial_0\tau}=\left( \bar{\tau}-\xi\bar{\lambda}\right) \gamma^0 .
\end{equation}
We therefore impose that
\begin{equation}
\left\lbrace \lambda_\alpha(x),\left( \bar{\lambda}(y)\gamma^0-\xi\bar{\tau}(y)\gamma^0\right)_\beta \right\rbrace=\left\lbrace \tau_\alpha(x),\left( \bar{\tau}(y)\gamma^0-\xi\bar{\lambda}(y)\gamma^0\right)_\beta \right\rbrace=\mathbbm{1}_{\alpha\beta}\delta^{(3)}\left(x-y\right) ,
\end{equation}
and put fermions to zero after calculating the anti-comutators.
We also set the gauge scalars to zero and impose the standard assumptions
for magnetic vortex solutions (the gauge choice $A_0=C_0=0$ and  time-independence of the
configuration). The resulting energy and central charge are,

\begin{eqnarray}
E &=&\int d^2x\left( \frac{1}{2}F_{ij}F^{ij}+\frac{1}{2}G_{ij}G^{ij}-\xi F^{ij}G_{ij}-D^2-d^2+\xi Dd+|D^As|^2+|D^Ct|^2\right) \nonumber\\
&=&\int d^2x\left(e \frac{1}{2}F_{ij}F^{ij}+\frac{1}{2}G_{ij}G^{ij}-\xi F^{ij}G_{ij}+V(s,t)+|D^As|^2+|D^Ct|^2\right),
\end{eqnarray}

\begin{equation}
T=i\epsilon^{ij}\int d^2x\left(\vphantom{\frac12} F_{ij}D+G_{ij}d-\xi F_{ij}d-\xi G_{ij}D+\left(D^A_is \right) \left(D^A_js \right)^\ast+ \left(D^C_it \right) \left(D^C_jt \right)^\ast\right).
\end{equation}
Inserting auxiliary fields $D$ and $d$ as given in eq.\eqref{DdDd}
we get
\begin{equation}
T=-\int d^2x\left(\epsilon^{ij}F_{ij}\frac{e}{2}\left(|s|^2-s_0^2\right)
+\epsilon^{ij}G_{ij}\frac{g}{2}\left(|t|^2-t_0^2\right)
+i\epsilon^{ij}\left(D^A_i s\right) \left(D^A_j s\right)^\ast  +i\epsilon^{ij}\left(D^C_i t \right) \left(D^C_j t \right)^\ast\right)
\end{equation}
which can be rewriten as
\begin{equation}
T=\int d^2x\text{  } \partial_i \left(\large\pi^i_1+\pi^i_2 \right),
\end{equation}
with
\begin{equation}
\pi^i_1=\epsilon^{ij}\left(A_j\frac{e}{2} s_0^2+is^\ast D^A_j s\right)\text{  ,  }\pi^i_2=\epsilon^{ij}\left(C_j\frac{g}{2} t_0^2+it^\ast D^C_j t\right).
\end{equation}\label{centralchargedensities}
Using Stoke's theorem we are left with a contour integral of these quantities over the circle at infinity. Since covariant derivatives should vanish on this contour, we get
\begin{equation}
T= {e}  s_0^2\oint A_idx^i + {g} t_0^2 \oint C_i dx^i =  s_0^2 {2\pi |n|}+
 t_0^2  {2\pi |k|}.
 \label{boundit2}
\end{equation}
Then, using eq.\eqref{desig} we obtain the same bound as in eq. 
\eqref{boundit}. We used the more
algorithmic procedure based on the SUSY algebra.

It should be noted that the interaction between vortices contributes no net central
charge (i.e. $T$ does not depend on $\xi$), only extra energy
through the gauge quadratic term
and the extra part of the scalar potential, as seen above.
This is natural; $T$ is a topological quantity. It cannot depend on
 smoothly-varying parameters.
\section{Numerical Solutions}\label{section5}

In this section we present the vortex solutions to eqs.(\ref{351})-(\ref{bps353}) obtained
using
an asymptotic shooting method \cite{art}. We use the Euler-Lagrange radial equations (3.42)-(3.45) since, being second order, there are two integration constants per equation in contrast with just one in the first order Bogomol'nyi case. This gives more degrees of freedom for the method to act on, and bases itself on a system that suffers less unstable behaviours as we move through parameter space, justifying the use of these (a priori more difficult) equations.

 Of course, we finally achieve complete agreement for the solutions of both first and second order systems. We plot the solution profiles and its dependence on the free  parameters $[\xi, e_r,\mu]$ defined in eq.(\ref{xxy}), showing that   these parameters can be tuned to have different profiles and widths for the hidden and visible strings. The idea is that this analysis could be useful to determine the best conditions and experimental framework to study the hidden sector, 
specially in connection with hidden dark strings interacting with
particles of the Standard Model.
We will  keep
the kinetic gauge mixing  $\xi$  in the region
$\xi < 10^{-3}$ since larger values are experimentally ruled out. Concerning  $e_r$, its value
  controls the vacuum expectation values  of the visible and hidden scalar fields.

We started our analysis by considering the $\xi\to 0 $ 
limit in which, as expected, the solutions correspond to those of 
two decoupled Abelian Higgs models, namely  the usual
Nielsen-Olesen vortices.

Then, by taking larger values of the mixing parameter we studied
the deviation of vortex solutions (both of the Higgs fields $s,t$ profiles 
and the magnetic. Below we present the most relevant features of the resulting solutions.

\begin{itemize}
\item {\it The decoupled case}: We first considered a small value of 
the kinetic mixing parameter, $\xi=10^{-8}$, and \underline{identical}  values for  
the gauge couplings and vector masses in both sectors, $e_r=\mu=1$. 
As advanced, the two sectors decouple showing each one Nielsen-Olesen vortex solutions..

Nonetheless, still in the very small  mixing parameter regime,  
apreciable departures from the decoupled Nielsen-Olesen solutions were found if, 
for instance, the two gauge coupling constant are different, $e_r\neq 1$. 
In this case, the rescaled VEVs of the two Abelian Higgs are different, 
see eqs.~(\ref{V1})-(\ref{V2}), 
therefore showing different profiles. 
The fields profile for the case $e_r=0.5$ is shown in Figure~(\ref{fig:fig1}). 
As can be seen from this figure, when $e_r$ decreases (for fixed $\mu$), 
the expectation value of the hidden Higgs field grows. 
Note that in this case the magnetic fields from both sectors remain identical.

We studied  larger values  $e_r>1$ and the result 
above still holds (the rescaled potential minimum of the hidden 
scalar is smaller). The main feature to be retained from these results  is that even in the very small $\xi$ regime, the Higgs scalars in each sector detect the gauge mixing showing different profiles while the magnetic fields remain indistinguishable when the gauge field masses are identical, see Figure~(\ref{fig:fig1}).

\begin{figure}[H]
\centering
\includegraphics[scale=0.75]{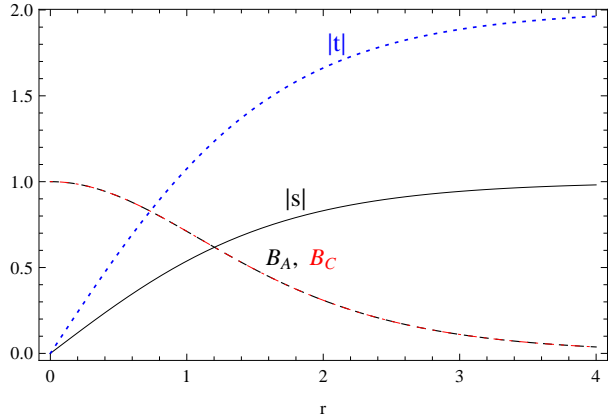}
\caption{Scalar and magnetic field profiles for $\xi=10^{-8}$,  $e_r=0.5$, $\mu=1$.}
\label{fig:fig1}
\end{figure}


It is important to note at this point that the actual parameter controlling the strength of the mixing in the kinetic sector is in fact given by the quotient $\xi/e_r$, as can be seen from equation (\ref{ener}). Then, even if in agreement with experimental 
constraints  one considers very small $\xi$  values,  the ratio $\xi/e_r$ can be made closer, or even bigger than one by considering the visible gauge coupling constant   much bigger than the  hidden sector one, this implying $e_r \ll 1$. Such possibility  is shown in Figure (\ref{fig:fig2}), in which  $\xi=10^{-8}$ and $e_r=10^{-8}$. One can see that in that case the visible and hidden 
magnetic fields differ. Concerning scalars,   the visible Higgs field profile remains similar to that in Figure~(\ref{fig:fig1}), while the  hidden Higgs one cannot be  shown in the plot becaused of its  much bigger VEV.

\begin{figure}[H]
\centering
\includegraphics[scale=0.75]{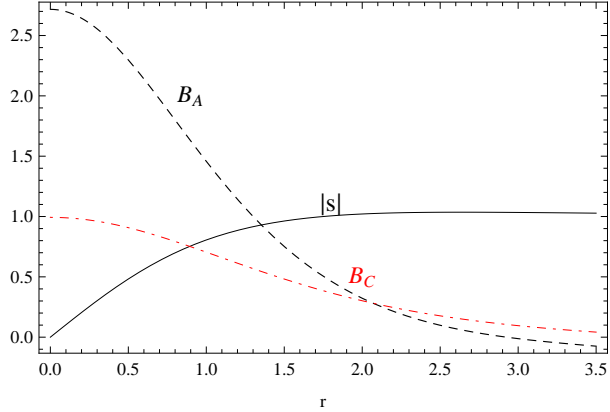}
\caption{Field profiles for  $\xi=10^{-8}$ and $e_r=10^{-8}$,  $\mu=1$. The hidden Higgs scalar $t$ is not shown in the figure because of the different scales involved (its VEV is too large in comparison with the visible one)}
\label{fig:fig2}
\end{figure}

Concerning the case in which the masses of the gauge fields 
are different $\mu\neq 1$ 
(with the mixing parameter   still very small, $\chi=10^{-8}$)  
none of the fields profile 
is identical to their hidden counterpart, as can be seen in Figure (\ref{fig:fig3}) 
for the case $\mu = 0.5$. Moreover, each sector exhibits profiles which coincide with the uncoupled
($\xi = 0$) case. Note that the choice
   corresponds to a hidden gauge field mass   smaller than the visible one this implying 
that  the exponential decay of the hidden magnetic field is slower, as can be seen in the figure. Now, since we are at the Bogomol'nyi point the mass of the hidden scalar  is identical to the hidden gauge field mass, and hence the growth of the scalar field towards its VEV is slower.

\begin{figure}[H]
\centering
\includegraphics[scale=0.75]{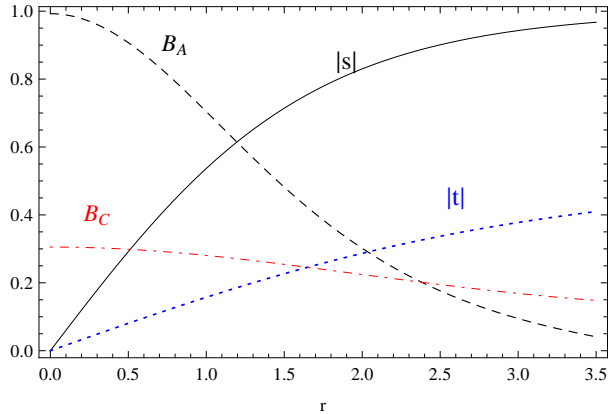}
\caption{We have considered $\mu=0.5$, $\xi=10^{-8}$ and $e_r=1$.}
\label{fig:fig3}
\end{figure}

\item{\it ``Strong'' mixing regime:} As explained in section 2.2 consistency of asymptotic behaviors implies that $\xi <1$. We then  consider that  the two sectors are strongly  coupled for the gauge mixing paramater of the order $\xi \sim 0.5$ taking  the ratio of gauge couplings  $e_r=0.5$. The profile of the fields is shown in Figure (\ref{fig:fig4}) for the case $\xi \sim 0.5$ so that  $\xi/e_r=1$. One can see that the visible and hidden magnetic field profiles  not only are different but the 
exhibit a crossover.
\end{itemize}

\begin{figure}[h]
\centering
\includegraphics[scale=0.75]{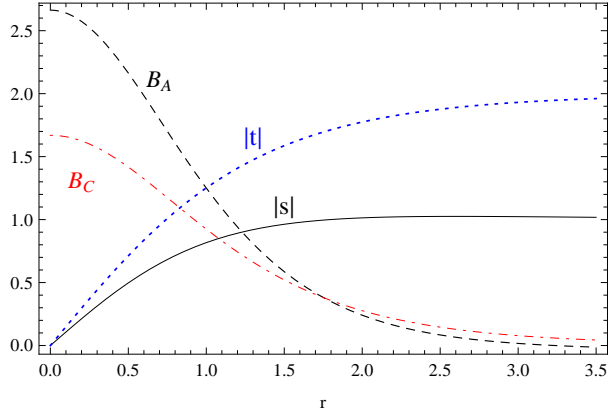}
\caption{We have considered $\mu=1$, $\xi=0.5$ and $e_r=0.5$.}
\label{fig:fig4}
\end{figure}


\section{Diagonalisation of the theory}\label{section6}
\subsection{The diagonal action}
{In his study of ``millicharged-particles''   \cite{Holdom},
Holdom observed that when the gauge fields are mixed with a $F_{\mu\nu}G^{\mu\nu}$ term,
one can perform a change of basis leading to an orthogonal diagonalisation
of the kinetic terms. In this section we extend such procedure to the
supersymmetric model extension that we are discussing.}

{We start from  the $N=2$ superspace Action in eq.\eqref{cici}
\begin{equation}
\begin{split}
S_{N=2}=\int d^3x d^2\theta d^2\bar{\theta} \lbrace \frac{1}{4}\Sigma\Sigma+\frac{1}{4}\Phi^\dagger e^{-ieU}\Phi +\frac{ies_0^2}{2}U+ \frac{1}{4}\Upsilon\Upsilon+\frac{1}{4}\Psi^\dagger e^{-igV}\Psi   +\frac{igt^2_0}{2}V -\frac{\xi}{2}\Sigma\Upsilon\rbrace.
\end{split}
\end{equation}
and, as we did in Section \ref{scalapotential}
for the scalar potential, we rewrite the supersymmetric Lagrangian
for the coupled  gauge fields sector   as a quadratic form;
\begin{eqnarray}
L_{\Sigma \Upsilon} &=& \frac14\left( \begin{array}{cc}
\Sigma & \Upsilon
\end{array} \right)
\left( \begin{array}{cc}
1 & -\xi \\
-\xi & 1
\end{array}\right)
\left( \begin{array}{c}
\Sigma \\
\Upsilon
\end{array} \right) \nonumber\\
 &=& \frac14 \left( \begin{array}{cc}
\Sigma & \Upsilon
\end{array} \right) {\cal M} \left( \begin{array}{c}
\Sigma \\
\Upsilon
\end{array} \right)
\label{arriba}.
\end{eqnarray}
The matrix ${\cal M}$ in the second line of eq. \eqref{arriba}
can be diagonalised by the following orthogonal matrix
\begin{equation}
\frac{1}{\sqrt{2}}\left( \begin{array}{cc}
1 & 1 \\
1& -1
\end{array}\right)
{\cal M}
\frac{1}{\sqrt{2}}\left( \begin{array}{cc}
1 & 1 \\
1& -1
\end{array}\right)=\left( \begin{array}{cc}
1-\xi & 0 \\
0& 1+\xi
\end{array}\right).
\end{equation}
Hence we define new decoupled gauge multiplets,
\begin{equation}
\left( \begin{array}{c}
\tilde{U}\\
\tilde{V}
\end{array} \right)=\frac{1}{\sqrt{2}}\left( \begin{array}{cc}
1 & 1 \\
1& -1
\end{array}\right)
\left( \begin{array}{c}
U \\
V
\end{array} \right).\label{diagonalisation}
\end{equation}}
{Note that this change of basis is its own inverse and it
induces a trivial Jacobian in the putative  partition function of the model.}

The new action, in terms of these gauge fields becomes,
\begin{eqnarray}
S_{N=2} &=&\int d^3x d^2\theta d^2\bar{\theta} \left( \frac{1-\xi}{4}\tilde{\Sigma}\tilde{\Sigma}+\frac{1}{2}\Phi^\dagger e^{-\frac{ie}{\sqrt{2}}(\tilde{U}+\tilde{V})}\Phi +\frac{ies^2_0}{2\sqrt{2}}\left( \tilde{U}+\tilde{V}\right)\right.\nonumber\\
&& + \left.\frac{1+\xi}{4}\tilde{\Upsilon}\tilde{\Upsilon}+\frac{1}{4}\Psi^\dagger e^{-\frac{ig}{\sqrt{2}}(\tilde{U}-\tilde{V})}\Psi   +\frac{igt^2_0}{2\sqrt{2}}\left( \tilde{U}-\tilde{V}\right) \right).
\end{eqnarray}
If $\xi=\pm 1$ one of these eigenvalues would vanish,
meaning one of the new gauge fields decouples. See below eq.(\ref{colores}) for an alternative view
on the same effect.
In order to canonically normalize the gauge kinetic terms we redefine superfields

\begin{equation}
\tilde{U}=\frac{1}{\sqrt{1-\xi}}\hat{U}\text{ , }\tilde{V}=\frac{1}{\sqrt{1+\xi}}\hat{V},\label{normalisation}
\end{equation}
and gauge charges,
\begin{equation}
e_1=\frac{e}{\sqrt{2(1-\xi)}}\text{ , }e_2=\frac{e}{\sqrt{2(1+\xi)}}\text{ , }g_1=\frac{g}{\sqrt{2(1-\xi)}}\text{ , }g_2=\frac{-g}{\sqrt{2(1+\xi)}},\label{diagonalcouplings}
\end{equation}
so that we end up with the following Action,

\begin{eqnarray}
S_{N=2} &=& \int d^3x d^2\theta d^2\bar{\theta} \left( \frac{1}{4}\hat{\Sigma}\hat{\Sigma}+\frac{1}{4}\Phi^\dagger e^{(-ie_1\hat{U}-ie_2\hat{V})}\Phi +\frac{1}{2}(ie_1s_0^2+ig_1t_0^2)\hat{U} \right.\nonumber \\ && + \left.\frac{1}{4}\hat{\Upsilon}\hat{\Upsilon}+\frac{1}{4}\Psi^\dagger e^{(-ig_1\hat{U}-ig_2\hat{V})}\Psi +\frac{1}{2}(ie_2s_0^2+ig_2t^2_0)\hat{V} \right).
\end{eqnarray}
This diagonal action is a more standard theory than the original one.
Indeed, the matter sector is in a bi-fundamental representation
of a $U(1)\times U(1)$ gauge group, with no direct interaction between
the gauge particles themselves. To recapitulate the new field content we have is,

\begin{equation}
\begin{split}
\hat{A}_\mu=\sqrt{\frac{1-\xi}{2}}(A_\mu+C_\mu)\text{ , } \hat{M}=\sqrt{\frac{1-\xi}{2}}(M+N)\text{ , }
\hat{\lambda}=\sqrt{\frac{1-\xi}{2}}(\lambda+\tau)\text{ , }
\hat{D}=\sqrt{\frac{1-\xi}{2}}(D+d)\text{ , }\\
\hat{C}_\mu=\sqrt{\frac{1+\xi}{2}}(A_\mu-C_\mu)\text{ , } \hat{M}=\sqrt{\frac{1+\xi}{2}}(M-N)\text{ , }
\hat{\tau}=\sqrt{\frac{1+\xi}{2}}(\lambda-\tau)\text{ , }
\hat{d}=\sqrt{\frac{1+\xi}{2}}(D-d).
\end{split}
\end{equation}
With these definitions, the Action now reads
\begin{eqnarray}
S &=&\int d^3x \left(-\frac{1}{4}\hat{F}_{\mu\nu}\hat{F}^{\mu\nu}-\half s^\dagger \acute{\Box}
s+\frac{1}{2}D^2-\left( \frac{ie_1}{2}\left(|s|^2-s_0^2\right)+\frac{ig_1}{2}\left(|t|^2-t_0^2\right)\right) \hat{D}
\right),\nonumber\\
&& -\frac{1}{2}\hat{M}\Box \hat{M}-\frac{1}{4}\hat{M}^2|s|^2+\frac{i}{2}\bar{\psi}\acute{\slashed{D}}\psi
+\frac{i}{2}\bar{\hat{\lambda}}\slashed{\partial}\hat{\lambda}-\frac{1}{2}\hat{M}\bar{\psi}\psi-\frac{e}{2}\left(\bar{\psi}\hat{\lambda} s+s^\dagger\bar{\hat{\lambda}}\psi\right),\nonumber \\
&& -\frac{1}{4}\hat{G}_{\mu\nu}\hat{G}^{\mu\nu}-\half t^\dagger \check{\Box} t+\frac{1}{2}d^2-\left( \frac{ie_2}{2}\left(|s|^2-s_0^2\right)+\frac{ig_2}{2}\left(|t|^2-t_0^2\right)\right)\hat{d}, \nonumber\\
&& -\left.\frac{1}{2}\hat{N}\Box\hat{N}-\frac{1}{4}\hat{N}^2|t|^2
+\frac{i}{2}\bar{\sigma}\check{\slashed{D}}\sigma+\frac{i}{2}\bar{\hat{\tau}}\slashed{\partial}\hat{\tau}-\frac{1}{2}\hat{N}\bar{\sigma}\sigma-\frac{g}{2}\left(\bar{\sigma}\hat{\tau} t+t^\dagger\bar{\hat{\tau}}\sigma\right),
\right)
\end{eqnarray}

where the covariant derivatives for the matter sector are now,

\begin{equation}
\acute{D}_\mu = (\partial_\mu -ie_1\hat{A}_\mu -ie_2\hat{C}_\mu )\;\;\; \text{ , } \;\;\;  \check{D}_\mu = (\partial_\mu -ig_1\hat{A}_\mu-ig_2\hat{C}_\mu).
\label{678}
\end{equation}

\subsection{Scalar potentials}
Since we have not redefined the scalar multiplets at all,
solving for the auxiliary fields and restoring the old gauge couplings
should give back the {previously obtained} potential.
Let us check this whilst also writing the scalar potential in terms of the new couplings.
The absence of the gauge mixing term means that the new auxiliary $D$-terms are
not intertwined and therefore can be eliminated independently;
\begin{align}
\hat{D}=\frac{ie_1}{2}\left(|s|^2-s_0^2 \right)+\frac{ig_1}{2}\left(|t|^2-t_0^2 \right) \\
\hat{d}=\frac{ie_2}{2}\left(|s|^2-s_0^2 \right)+\frac{ig_2}{2}\left( |t|^2-t_0^2\right).
\end{align}
which,
after substitution for the couplings given by eq.(\ref{diagonalcouplings}) gives
rise to the following potential,

\begin{eqnarray}
V\left[s,t \right] &=&\left(\frac{ e_1^2 +e_2^2}{8} \left( |s|^2-s_0^2\right)^2+\frac{ g_1^2 +g_2^2}{8} \left( |t|^2-t_0^2\right)^2+\frac{e_1g_1+e_2g_2}{4}\left(  |s|^2-s_0^2\right) \left(|t|^2-t_0^2 \right)  \right) \nonumber\\
&=&\frac{1}{1-\xi^2}\left(\frac{e^2}{8} \left( |s|^2-s_0^2\right)^2+\frac{ g^2}{8} \left( |t|^2-t_0^2\right)^2+\frac{eg\xi}{4}\left(  |s|^2-s_0^2\right) \left(|t|^2-t_0^2 \right)  \right).
\end{eqnarray}
As expected this expression coincides with the previously obtained
scalar potential eq. \eqref{cahpotential}.

\subsection{Supercharges and Algebra}
The diagonalisation  allows us to easily find the Bogomol'nyi bound for the energy from the
simpler supercharge algebra.
Indeed, the diagonalized  Lagrangian possesses the following supercharges,
\begin{eqnarray}
Q_{S} &=&\int d^2x \left( \left( -\frac{1}{2} \epsilon^{\mu\nu\rho}\hat{F}_{\mu\nu} \gamma_\rho+i\slashed{\partial}\hat{M}+i\hat{D}^\ast \right)\gamma^0 \hat{\lambda} +\left( i\left( \slashed{\partial}-ie_1\slashed{\hat{A}}-ie_2\slashed{\hat{C}}\right)s^{\ast}-\frac{e_1}{2}\hat{M}s^\ast -\frac{e_2}{2}\hat{N}s^\ast \right)\gamma^0 \psi \right.\nonumber\\
&& + \left.
\left( -\frac{1}{2} \epsilon^{\mu\nu\rho}\hat{G}_{\mu\nu} \gamma_\rho+i\slashed{\partial}\hat{N}+i\hat{d}^\ast \right)\gamma^0 \hat{\tau}
+\left( i\left( \slashed{\partial}+ig_1\slashed{\hat{A}}+ig_2\slashed{\hat{C}}\right)t^{\ast}-\frac{g_1}{2}Mt^\ast -\frac{g_2}{2}Nt^\ast\right)\gamma^0 \sigma \rbrace
\right),
\end{eqnarray}
and
\begin{eqnarray}
\bar{Q}_{S}&=&\int d^2x \left( \bar{\hat{\lambda}}\gamma^0 \left( -\frac{1}{2} \epsilon^{\mu\nu\rho}\hat{F}_{\mu\nu} \gamma_\rho-i\slashed{\partial}\hat{M}-i\hat{D} \right)
 +\bar{\psi}\gamma^0\left( -i\left( \slashed{\partial}-ie_1\slashed{\hat{A}}-ie_2\slashed{\hat{C}}\right)s-\frac{e_1}{2}\hat{M}s -\frac{e_2}{2}\hat{N}s \right)
\right. \nonumber\\ &&
\left.\bar{\hat{\tau}}\gamma^0 \left( -\frac{1}{2} \epsilon^{\mu\nu\rho}\hat{G}_{\mu\nu} \gamma_\rho-i\slashed{\partial}\hat{N}-i\hat{d} \right)+\bar{\sigma}\gamma^0\left(\left( \slashed{\partial}-ig_1\slashed{\hat{A}}-ig_2\slashed{\hat{C}}\right)t-\frac{g_1}{2}Mt -\frac{g_2}{2}Nt\right)  \right).
\end{eqnarray}
Note that in this case the conjugate momenta of the fermions are the canonical
ones so the procedure is simpler. Calculating
the SUSY algebra and imposing that the fermions and gauge
scalars vanish, we are left with the following energy and central charge,

\begin{equation}
E=\int d^2x \lbrace \frac{1}{2}\hat{F}_{ij}\hat{F}^{ij}+|\acute{D}s|^2+\frac{1}{2}\hat{G}_{ij}\hat{G}^{ij}+|\check{D} t|^2+V(s,t)\rbrace.
\end{equation}

\begin{eqnarray}
T &=&-\int d^2x\left( \epsilon_{ij}\hat{F}^{ij}\left( \frac{e_1}{2}\left(|s|^2-s_0^2 \right)+\frac{g_1}{2}\left( |t|^2-t_0^2\right) \right)  +i\epsilon^{ij}\left(\acute{D}_is\right)\left(\acute{D}_js  \right)^\ast \right. \nonumber \\
&& \left.
+\,\epsilon_{ij}\hat{G}^{ij}\left(\frac{e_2}{2}\left(|s|^2-s_0^2\right) +\frac{g_2}{2}\left( |t|^2-t_0^2\right) \right)
 +i \epsilon^{ij}\left(\check{D}_it\right)\left(\check{D}_jt  \right)^\ast \right).
\end{eqnarray}
As before  this central charge can be written as a total derivative,
\begin{equation}
T=\int d^2x \text{   } \partial_i\mathcal{V}^i
\end{equation}
with
\begin{eqnarray}
 \mathcal{V}^i &=& \epsilon^{ij}\left(\hat{A}_j\left(\frac{e_1}{2}s_0^2+\frac{g_1}{2}t_0^2 \right)+\frac{i}{2}s^\ast\left(\partial_j-ie_1\hat{A}_j-ie_2\hat{C}_j \right) s \right. \nonumber \\
&& \left.+\,\hat{C}_j\left(\frac{e_2}{2}s_0^2+\frac{g_2}{2}t_0^2 \right)+\frac{i}{2}t^\ast\left(\partial_j-ig_1\hat{A}_j-ig_2\hat{C}_j \right) t \right).
\end{eqnarray}
Then, using Stoke's theorem and imposing that covariant derivatives vanish at infinity in order to have finite energy, we get for the central charge
\begin{equation}
T=\oint dx^i\left( \left( e_1s_0^2+g_1t_0^2 \right)\hat{A}_i+ \left(e_2s_0^2+g_2t_0^2 \right)\hat{C}_i\right)=\left( e_1s_0^2+g_1t_0^2 \right)\Phi_{\hat{A}}+\left(e_2s_0^2+g_2t_0^2 \right)\Phi_{\hat{C}}.
\end{equation}

The energy is thus bounded by a linear combination of the fluxes (proportional to vortex units of flux  numbers) of these mixed gauge fields $\hat{A},\hat{C}$. We can restore the dependence on the original fields,
\begin{equation}
T=\oint dx^i\left( \left(\frac{e}{2}s_0^2+\frac{g}{2}t_0^2 \right)\left( A_i+C_i\right)+ \left(\frac{e}{2}s_0^2-\frac{g}{2}t_0^2 \right)\left( A_i-C_i\right)\right)=es_0^2\Phi_A+gt_0^2\Phi_C,
\end{equation}
which is precisely what we found earlier---see eqs.(\ref{boundit}) and (\ref{boundit2}).
For completeness, let us also restore the original fields in the energy,
\begin{eqnarray}
E &=&  \int d^2x \left( \frac{1}{4}(1-\xi)\left(F_{ij}+G^{ij}\right)^2+\frac{1}{4}(1+\xi)\left(F_{ij}-G^{ij} \right) +V(s,t)+|D_i^A s|^2+|D_j^Ct|^2
\right.\nonumber\\
&=& \left.\int d^2x\lbrace \frac{1}{2}F_{ij}F^{ij}+\frac{1}{2}G_{ij}G^{ij}-\xi F^{ij}G_{ij}+V(s,t)+|D_i^As|^2+|D_i^Ct|^2\right).
\end{eqnarray}
Again this is perfectly consistent with the results obtained previously.
It is worth noting that the standard Bogomol'nyi approach to finding these
quantities by completing various positive terms in the action is,
in this circumstance, far more obvious, given that there is no gauge mixing terms.
It is indeed, just a question of completing some squares.
The positive definite quadratic form that appeared in our previous
theory has been diagonalised away.

\subsection{Applications of the diagonal theory: correlation functions}
Here, in a somewhat unrelated development, we consider a nice implication for the QFT
aspects of the gauge kinetic mixed theory as read from the diagonalised theory.

{The two theories are exactly equivalent, even at the quantum level
since the Jacobian of our transformation is trivial.
If we add currents to the partition function, via a term}
\begin{equation}
\int d^3x \left(J^A_\mu A^\mu+J^C_\mu C^\nu\right)=\int d^3x \left(  \begin{array}{cc}
J^A_\mu & J^C\nu
\end{array} \right)
\left( \begin{array}{c}
A^\mu \\
C^\nu
 \end{array} \right).
\end{equation}
We retain the shape of this form after diagonalisation as long
as we perform the \textit{opposite} transformation on the currents;
\begin{equation}
\left( \begin{array}{c}
J^{A}_\mu \\
J^{C}_\nu
\end{array}\right)  = \left(\begin{array}{cc}
\sqrt{1-\xi} & 0 \\
0 & \sqrt{1+\xi}
\end{array}  \right)\frac{1}{\sqrt{2}}\left( \begin{array}{cc}
1 & 1 \\
1 & -1
\end{array}  \right) \left( \begin{array}{c}
J^{\hat{A}}_\mu \\
J^{\hat{C}}_\nu
\end{array} \right).
\end{equation}
Thus we can write
\be
\dfrac{\delta}{\delta J^{C}_\mu}=\dfrac{1}{\sqrt{2}}\left(\dfrac{1}{\sqrt{1-\xi}}
\dfrac{\delta}{\delta J^{\hat{A}}_\mu}-\dfrac{1}{\sqrt{1+\xi}}
\dfrac{\delta}{\delta J^{\hat{C}}_\mu}\right),
\ee
and similarly for the other transformed current. This gives an
explicit formula to transform correlation functions in one theory to
those in the other theory, thus, we can calculate every observable of one theory
from observables of the other.

For instance, {looking at the mixing  terms diagrammatically},
allows us to add a 2-point vertex transforming $A$ into $C$
with amplitude $\xi$. This is analytically similar to a mass term,
in that an arbitrary amount of this vertex can be added to any gauge propagator,
which then need to be summed over as a geometric series.
In practice: For the $A-A$ and $C-C$ propagator (with the same ingoing and outgoing particle states), one can add any even number of this interaction 2-vertex, leading to a factor $\xi^{2n}$ for each of them, summing over them means that these propagators get modified by a factor of $ {1}/(1-\xi^2)$. In addition, the Feynman rules now also possess a $A-C$ propagator with different in and out states, corresponding to an odd number of inserted vertices, thus it has a factor of $ {\xi}/(1-\xi^2)$. This allows for extra channels: one can turn a pair of scalars of one sector into the other scalar pair using this mixed-states propagator.

This is consistent with the results from the diagonalised theory. With these modified gauge propagators, the $ss\rightarrow tt$ tree-level amplitude is proportional to $ {eg\xi}/(1-\xi^2)$, in the diagonalised theory, summing over both channels we get
\begin{equation}
\frac{eg}{2}\left(\frac{1}{1-\xi}-\frac{1}{1+\xi} \right)= \frac{eg\xi}{1-\xi^2}.
\end{equation}
From this point of view, neither scalar gains an effective charge under the other gauge field. Rather, the gauge field oscillates as it propagates and allows for production of particles in the other sector. Had we performed a non-orthogonal change of basis, such as

\begin{equation}
\tilde{A}=A-\xi C\text{  ,  }\tilde{C}=\sqrt{1-\xi^2 }C,
\end{equation}
then $s$ gains (in terms of the new gauge fields) a small hidden sector charge
{${e\xi}/{\sqrt{1-\xi^2}}$, while the charge of $t$ is rescaled
to ${g}/{\sqrt{1-\xi^2}}$} leading to the same $ss\rightarrow tt$
tree-level amplitude. This approach, while consistent, artificially breaks the
equivalence of the two sectors. Indeed, the content of both sectors have
the same structure, and are made to communicate by a term that is $A\leftrightarrow C$ invariant, it is more elegant to find an interpretation of this term (i.e. a reformulation of the theory) that does not disturb this property.

Let us now summarize our findings and propose some topics for future investigation.

\section{Conclusions}\label{section7}
In this paper we have been partly motivated by models for 
the hidden sector of different
mechanisms of SUSY breaking
in beyond the Standard Model Physics and also by models with a Higgs portal
and gauge kinetic mixing interaction. But mostly our motivation came from recent
developments on Dark Matter and topics around that. Indeed, in some scenarios, the  dark-sector
is modelled by a lagrangian that communicates it with the Standard Model via
a gauge-kinetic mixing interaction. The problem that occupied us in this work
was the study of dark-strings, namely topological defects of the dark sector,
when in interaction with the visible sector via terms discussed above. 
Due to different observational constraints, we considered the situation in which both $U(1)$'s---
the Standard Model and the hidden one---are spontaneously broken. At large distances
this system is well described by two  Abelian Higgs models that interact via
a gauge kinetic mixing term and a potential to be determined.

We searched for topological objects  in this model, using a well established
procedure; namely we extended the model of eq.(\ref{actione}) to ${\cal N}=2$ SUSY.
We then read BPS equations and topological charges using the SUSY algebra. We have
checked our results with more traditional methods, finding complete agreement.

In fact, the ${\cal N}=2$
version of the model of \cite{AS} determines the interaction potential and
relations between different couplings, so that the model presents stable vortex solutions,
generalising those of Nielsen-Olesen. As mentioned, we found the topological charge that
bounds the Energy of our strings and BPS equations that control the string dynamics.
We have studied these equations numerically, finding the set of parameters that control the
shapes and widths of the hidden and visible strings. 
The relevant parameters are: $\mu$ controling the quotient
of the masses of the (spontaneously broken) gauge fields; $e_r$ that controls
 the VEV of the hidden Higgs field $t$. Finally, the parameter $\frac{\xi}{e_r}$,  accounts
for the strength of the interaction.

We observed that in the 'decoupled' case $\frac{\xi}{e_r}\to 0$, 
for equal masses of 
the gauge fields $\mu=1$ and fixed VEV for the hidden Higgs 
($e_r\sim 1$), we are in 
the expected situation of a pair of decoupled Abelian Higgs Models. 
Changing to $e_r\neq 1$,
we found departures from the fully decoupled case; 
the VEV $<t>$ is inversely proportional to
$e_r$, while the profiles of the magnetic fields are still similar.  
If the masses of the gauge fields
are taken to be different (for example $\mu<1$), 
the hidden and visible magnetic fields
decay differently, the hidden vortex is more 
delocalised (for $\mu<1$ and $e_r=1$). 
On the other hand, when we consider
a 'strongly mixed' situation, $\frac{\xi}{e_r}\sim 1$, the profiles of both
strings are different, even when the gauge fields acquire the same mass. 
See Figures~(\ref{fig:fig1})-(\ref{fig:fig4}) for an 
illustration of these points.

Finally, we closed our
study with a nice alternative way of obtaining these results, by considering a diagonal
basis of gauge fields (that on the other hand, charges both hidden and visible matter under
both the Standard Model and the hidden gauge groups). Indeed, the gauge kinetic mixing term's effects is to make the gauge fields oscillate from one sector to another during propagation, this diagonalisation argument is nothing more than a propagation eigenbasis for the theory. Of course, we obtained perfect agreement
using different perspectives. It is especially nice that differing formalisms
manage to produce a topological (central) charge that is---reasonably--- independent
of the parameter $\xi$ weighting the gauge kinetic mixing term.

Various problems for future study are suggested by the contents of this paper.
Given that the symmetries of the problem reduce it to three
space-time dimensions it seems natural  to study the behavior when a
 Chern-Simons term is present --see \cite{HKP}-\cite{ShifmanYung}-- for a sample of different
aspects of Chern-Simons SUSY Actions and vortex solutions.
The study of scattering of visible particles
with our topological strings is needed to make concrete predictions on cross sections,
that might be experimentally verified. In this sense a more robust numerical analysis
than the one presented here would be desirable.

It would also be interesting to study the extension of our formalism, in the case
in which the models are non-Abelian (hence applicable when the low Energy
description in this paper breaks down)--see \cite{Man} for some work on that direction.

We hope to have tempted
phenomenologically-minded readers to tackle some of this questions.


\section*{Acknowledgments:} It is a pleasure to thank Prem Kumar
 for discussions
and comments on the material presented in this 
paper and Enrique Moreno for comments and help in the numerical analysis. We also thank Daniel Thompson for his insight.

P. Arias is supported by FONDECYT project 11121403 and Anillo ACT 1102.
E. Ireson is supported by a STFC studentship. 
C. Nunez is Wolfson Fellow of the Royal Society. C. Nunez thanks IHES (France)
for hospitality.
F.A. Schaposnik is financially supported by CONICET, 
ANPCyT, UNLP and CICBA grants.

\begin{appendices}
\section{Derivatives in curvilinear coordinates}

In the context of using the Nielsen-Olesen Ansatz, we make use of the form of the Del operator $\nabla$ in curvilinear (cylindrical) coordinates. The reader may find it instructive to have a brief summary of how its exact form is derived.

In a generic system of orthogonal curvilinear coordinates, we can write the line element (without the summation convention)

\begin{equation}
ds^2=\sum_{\mu}(h^\mu)^2 (dx^\mu)^2
\end{equation}
The $h^\mu$ are called scale factors for the coordinate system. We perform a rescaling of our basis so as to make them orthonormal: define
\begin{equation}
e^{(\mu)}=h^\mu dx^\mu
\end{equation}
so that
\begin{equation}
ds^2=\sum_{\mu} (e^{(\mu)})^2
\end{equation}
It is in this system that it we define the field strength for the gauge field. Write the latter in the new coordinates:
\begin{equation}
A=\sum_{\mu} A_\mu dx^\mu=\sum_{\mu}A_{(\mu)}e^\mu=\sum_{\mu} A_{(\mu)}h^\mu dx^\mu,
\end{equation}
this gives the general change of basis formula
\begin{equation}
A_\mu=h^\mu A_{(\mu)}
\end{equation}
Henceforth we use the convention that bracketed indices express components of the tensor in the orthonormal basis, which we can then re-expressed  using the previous relation in the original basis. To find the field strength, we take the (exterior) derivative of this object: in our original curvilinear coordinates, we get components of $\nabla_\mu A_\nu$, which we want to find. Altogether,
\begin{equation}
dA \equiv\sum_{\mu\nu}(\nabla_\mu A_\nu)dx^\mu\wedge dx^\nu=\sum_{\mu\nu}(\partial_\mu A_{(\nu)})e^\mu\wedge e^\nu.
\end{equation}
The derivative on the right hand side of this equation is the "standard" derivative of a function with respect to the coordinate, the basis $e^\mu$ encapsulates all the data relative to the geometry of the space. This is not the case in the original coordinate basis.

Let as now make a change of basis
\begin{equation}
dA=\sum_{\mu\nu}\frac{1}{h^\mu h^\nu}\partial_\mu (h^\nu A_\nu) dx^\mu \wedge dx^\nu.
\end{equation}
We can now identify the unknown quantity $\nabla A$ component-wise with the components of this expression and get
\begin{equation}
\nabla_\mu A_\nu=\frac{1}{h^\mu h^\nu}\partial_\mu (h^\nu A_\nu).
\end{equation}
This the required formula. In practice for cylindrical coordinates: we have two components labelled $r,\theta$ of scale factors
\begin{equation}
h^r=1\text{ , }h^\theta=r,
\end{equation}
then, this implies that
\begin{equation}
\nabla_r A_\theta=\frac{1}{  r}\partial_r \left(rA_\theta \right)
\end{equation}
which is what we used in our equations of motion.

This is of course consistent with the more general formalism of General Relativity, but this method circumvents setting up the correct framework (tangent and co-tangent bundles) and having to calculate Christoffel symbols. It also makes good use of the particular, diagonal form of the metric in a curvilinear coordinate system.

\end{appendices}


\begin{thebibliography}{99}
  \bibitem{SZ}
  V.~Silveira and A.~Zee,
  Phys.\ Lett.\ B {\bf 161} (1985) 136.
\bibitem{PW}
  B.~Patt and F.~Wilczek,
  hep-ph/0605188.
  \bibitem{DKM}
  K.~R.~Dienes, C.~F.~Kolda and J.~March-Russell,
  Nucl.\ Phys.\ B {\bf 492} (1997) 104
  [hep-ph/9610479].
\bibitem{Okun} L.~B.~Okun,
  Sov.\ Phys.\ JETP {\bf 56} (1982) 502
   [Zh.\ Eksp.\ Teor.\ Fiz.\  {\bf 83} (1982) 892].
  \bibitem{Galison:1983pa}
  P.~Galison and A.~Manohar,
  Phys.\ Lett.\ B {\bf 136} (1984) 279.
\bibitem{Holdom}
  B.~Holdom,
  Phys.\ Lett.\ B {\bf 166} (1986) 196.
  \bibitem{Arkani}
  N.~Arkani-Hamed, D.~P.~Finkbeiner, T.~R.~Slatyer and N.~Weiner,
  ``A Theory of Dark Matter,''
  Phys.\ Rev.\ D {\bf 79} (2009) 015014
  [arXiv:0810.0713 [hep-ph]].
\bibitem{Davidson:2000hf}
  S.~Davidson, S.~Hannestad and G.~Raffelt,
  JHEP {\bf 0005}, 003 (2000)
  [hep-ph/0001179].

\bibitem{JR1}
  J.~Jaeckel and A.~Ringwald,
  Ann.\ Rev.\ Nucl.\ Part.\ Sci.\  {\bf 60} (2010) 405
  [arXiv:1002.0329 [hep-ph]].
\bibitem{Vachaspati}
  T.~Vachaspati,
  Phys.\ Rev.\ D {\bf 80} (2009) 063502
  [arXiv:0902.1764 [hep-ph]].
  \bibitem{betti}
  B.~Hartmann and F.~Arbabzadah,
  JHEP {\bf 0907}, 068 (2009)
  [arXiv:0904.4591 [hep-th]].
  \bibitem{Brihaye:2009fs}
  Y.~Brihaye and B.~Hartmann,
  Phys.\ Rev.\ D {\bf 80} (2009) 123502
  [arXiv:0907.3233 [hep-th]].
  \bibitem{Hyde}
  J.~M.~Hyde, A.~J.~Long and T.~Vachaspati,
  Phys.\ Rev.\ D {\bf 89} (2014) 065031
  [arXiv:1312.4573 [hep-ph]].
\bibitem{Nelson:2011sf}
 A.~E.~Nelson and J.~Scholtz,
  Phys.\ Rev.\ D {\bf 84} (2011) 103501
  [arXiv:1105.2812 [hep-ph]];
  P.~Arias, D.~Cadamuro, M.~Goodsell, J.~Jaeckel, J.~Redondo and A.~Ringwald,
  JCAP {\bf 1206} (2012) 013
  [arXiv:1201.5902 [hep-ph]].
  \bibitem{M2}
  D.~E.~Morrissey and A.~P.~Spray,
  JHEP {\bf 1406} (2014) 083
  [arXiv:1402.4817 [hep-ph]].
\bibitem{AS}
  P.~Arias and F.~A.~Schaposnik,
  arXiv:1407.2634 [hep-th].
  \bibitem{Bogo}
  E.~B.~Bogomol'nyi,
  Sov.\ J.\ Nucl.\ Phys.\  {\bf 24} (1976) 449
   [Yad.\ Fiz.\  {\bf 24} (1976) 861]. [Reprinted in
{\em Solitons and Particles}, Eds. C. Rebbi and G. Soliani (World
Scientific, Singapore, 1984), p. 389].
\bibitem{dVS}
  H.~J.~de Vega and F.~A.~Schaposnik,
  Phys.\ Rev.\ D {\bf 14} (1976) 1100. [Reprinted in
{\em Solitons and Particles}, Eds. C. Rebbi and G. Soliani (World
Scientific, Singapore, 1984), p. 389].
  \bibitem{WO}
  E.~Witten and D.~I.~Olive,
  Phys.\ Lett.\ B {\bf 78} (1978) 97.
  \bibitem{ENS1}
  J.~D.~Edelstein, C.~Nunez and F.~Schaposnik,
  ``Supersymmetry and Bogomolny equations in the Abelian Higgs model,''
  Phys.\ Lett.\ B {\bf 329} (1994) 39
  [hep-th/9311055].
  \bibitem{ENS2}
  J.~D.~Edelstein, C.~Nunez and F.~A.~Schaposnik,
  Nucl.\ Phys.\ B {\bf 458} (1996) 165
  [hep-th/9506147].
\bibitem{ENS3}
  J.~D.~Edelstein, C.~Nunez and F.~A.~Schaposnik,
  Phys.\ Lett.\ B {\bf 375}, 163 (1996)
  [hep-th/9512117].
\bibitem{art}  W.~H.~Press, S.~A.~Teukolsky, W.~V.~Vetterlink, {\it Numerical Recipes: The art of Scientific
Computing}, Cambridge University Press, Cambridge U.K. (1992).
\bibitem{HKP} J.~Hong, Y.~Kim, and P.~Y.~Pac, Phys.\ Rev.\ Lett.\ {\bf 64} (1990) 223
\bibitem{JW}
  R.~Jackiw and E.~J.~Weinberg,
  Phys.\ Rev.\ Lett.\  {\bf 64} (1990) 2234.
  \bibitem{Lee:1990it}
  C.~k.~Lee, K.~M.~Lee and E.~J.~Weinberg,
  Phys.\ Lett.\ B {\bf 243} (1990) 105.
%
\bibitem{Man} L.~F.~Cugliandolo, G.~Lozano, M.~V.~Manias and F.~A.~Schaposnik,
  Mod.\ Phys.\ Lett.\ A {\bf 6} (1991) 479.\bibitem{Dunne:1998qy}
  G.~V.~Dunne,
  {\it Aspects of Chern-Simons theory},
  hep-th/9902115.
\bibitem{Schwarz:2004yj}
J.~H.~Schwarz, ``Superconformal Chern-Simons theories,''
  JHEP {\bf 0411}, 078 (2004)  [hep-th/0411077].
\bibitem{Schaposnik:2006xt}
  F.~A.~Schaposnik, {\it Vortices},  hep-th/0611028.
\bibitem{ShifmanYung}   M.~Shifman and A.~Yung,
  {\it Supersymmetric solitons},
  Cambridge, UK: Cambridge Univ. Pr. (2009) 259 p
\end{thebibliography}
\end{document}